\documentstyle [12pt]{article}

\begin{document}

\title{ New Type of Fields in Maxwell's Equations: Their Properties and Method of
Detection}

\author{\bf A. G. Chirkov* and A. N. Ageev**}

\begin{center}

* St. Petersburg State Technical University, St.
Petersburg,195251 Russia; \ ** Ioffe Physicotechnical Institute,
Russian Academy of Sciences, St. Petersburg, 194021 Russia.
\end{center}

\date{\today}

\begin{abstract}

A theoretical analysis of the excitation of an infinitely long
solenoid by oscillating current has revealed the existence of
specific potentials in the space outside the solenoid, which can
affect electron diffraction in an experiment similar to the
Aharonov-Bohm effect. Thus, these time-dependent potentials are
physical fields possessing a number of specific features, which
set them off from the fields known heretofore.

\end{abstract}

\maketitle

The peculiar phenomenon, predicted in 1939 and 1949 [1, 2] and rediscovered
and studied theoretically in considerable detail in 1959 [3], was
subsequently called the Aharonov-Bohm (AB) effect. It consists essentially
in that in propagating through a region with no magnetic or electric field
present, but where the vector or scalar potential is nonzero, the de Broglie
wave corresponding to a quantum charged particle is acted upon by the
latter. These conditions are best realized in a static regime, which was
exactly the case studied before the 1990s. While a long discussion has
certainly contributed to a proper understanding of the AB effect (see, e.g.,
reviews [4, 5]), heated debates on this issue are still continuing in the
literature.

Based on the totality of the experiments performed, one has to admit that
the AB effect can exist only if there are potentials, which do not generate
fields and cannot be removed by gauge transformation. We have termed them
``zero-field potentials''.{\large} Note that zero-field potentials, which
transform only the phase of a wave function, are responsible for the AB
effect in all the papers published heretofore and dealing with the static
case. In a general case, such potentials satisfy the relations

 $ - \,c^{ - 1}\partial {\rm {\bf A}}^{0} / \partial t\, - \,grad\,\varphi
^{0}\,\,\, = \,\,0$ {\large and} $\,rot\,{\rm {\bf A}}^{0}\, = \,\,0,${\large
(1)}

\noindent
where the upper indices of the potentials refer to the zero-field
potentials. Because such potentials should obviously have the form

\begin{equation}
\label{eq1}
\,{\rm {\bf A}}^{0} = \,\,\,grad\,\chi {\rm è}
\quad
\varphi ^{0}\, = \, - \,{\frac{{1}}{{c}}}\,{\frac{{\partial \,\chi
}}{{\partial \,t}}},
\end{equation}

\noindent
the $\,\chi $ function was erroneously identified in practically all
publications with the gradient potential transformation function, and this
is what gives rise frequently to misunderstanding. In the case where $\chi $
is a differentiable function of its arguments and unique, Eq. (\ref{eq1}) defines
the gauge transformation of the potentials, ${\rm {\bf A}}\,\, \Rightarrow
\,{\rm {\bf A}}\prime $, which must satisfy two equivalence conditions

{\large 1)} local (differential), $\,rot\,{\rm {\bf A}}\,\, = \,\,rot\,{\rm
{\bf A}}\prime \,${\large ,} {\large (2a)}

{\large 2)} global (integral),{\large} ${\oint\limits_{L} {\,{\rm {\bf
A}}\,\,d{\rm {\bf l}}}} \,\, = \,{\oint\limits_{L} {\,{\rm {\bf A}}\prime
\,\,d{\rm {\bf l}}}} ${\large , (2b)}

As follows from Eq. {\large (2a)}, $\,{\rm {\bf A}}\prime = \,\,\,{\rm {\bf
A}}\,\, + \,\,\,grad\,\chi $. However, as is well known from the theory of
cohomology [6], in a general case ${\oint\limits_{L} {\,grad\,\chi \,\,d{\rm
{\bf l}}}} \,\, \ne \,\,0\,\,\,$. Obviously enough, in these conditions the
$\chi $ function must possess certain features, for instance, be
multivalued.

The papers dealing with the AB effect did not practically discuss the
necessary condition of its existence. In the static case, one usually
restricted oneself to maintaining that in order for the AB effect to exist,
the region of space where the quantum particle propagates should have a
nontrivial topology [3, 7]. This is, however, at odds with the experiment of
Chambers [8]. Indeed, in the case of an infinitely long solenoid or torus
the space is doubly connected, while in the experiment of Chambers, where a
magnetized filamentary iron single crystal of a finite length was used, the
space was singly-connected. As evident from many other experiments (see,
e.g., [4]), the AB effect, rather than depending on space topology, is
determined by the presence of zero-field potentials.

After the convincing experiments of Tonomura {\it et al}. [9], the possibilities of
studying the static AB effect at the present level of technology were
apparently exhausted, and the researchers turned their attention to the
investigation of the time-dependent, or quasi-AB effect [10].{\large
}However, in this work, which has certainly produced fruitful results, the
potentials responsible for the quasi-AB effect were introduced artificially,
without discussing in any way their nature. Nevertheless, the origin of
these potentials (fields) is a major issue in the separation of the AB
effect from the general variation of the de Broglie wave-interference
pattern.

We maintain that in the regions of space with no currents present the total
potentials can be presented, generally speaking, in the form

 ${\rm {\bf A}} = \,\,{\rm {\bf A}}^{f} + \,{\rm {\bf A}}^{0}$ and {\large
}$\varphi \, = \,\varphi ^{f}\, + \,\varphi ^{0}${\large , (3)}

\noindent
where index $f$ refers to ``field'' potentials corresponding to nonzero
electromagnetic fields:

\begin{equation}
\label{eq2}
{\rm {\bf E}} = \, - \,c^{ - 1}\partial {\rm {\bf A}}^{f} / \partial t\, -
\,grad\,\varphi ^{f}\,,
\quad
\,
{\rm {\bf B}} = \,rot\,{\rm {\bf A}}^{f}.
\end{equation}

Index $0$ in Eq. (3) identifies zero-field or excess potentials defined by
relations (1).{\large} Note that the ``excess'' potentials have been long in
use in mathematical physics [11]; they are necessary when solving Maxwell's
equations with boundary conditions. Thus, while for the field potentials
local quantities ${\rm {\bf E}}$ and ${\rm {\bf B}}$ of Eq. (\ref{eq2}) have a
physical (gauge-invariant) meaning, for the zero-field potentials this
meaning is found in integral quantities

\begin{equation}
\label{eq3}
\omega _{1} \, = \,{\oint\limits_{L} {{\rm {\bf A}}}} \,d\,{\rm {\bf l}}
\end{equation}

\noindent
in the static case, or

\begin{equation}
\label{eq4}
\omega _{2} \,\, = \,\,{\oint\limits_{S} {A^{i}\,d\,x_{i}}}  ,
\end{equation}

\noindent
in the time-dependent conditions, where $A^{i}\, = \,({\rm {\bf
A}},\,?\,\varphi ),\, \quad dx_{i} \, = \,\,(d{\rm {\bf r}},\,c\,dt)\,$, and the
integral in Eq. (\ref{eq4}) is calculated over a time-like surface.

We are going to demonstrate the above in a specific example.{\large
}Consider circular currents flowing in a region of space to form an
infinitely long cylinder of radius $R$ (a solenoid with circular currents).
Choose a cylindrical reference frame ($\rho ,\,\alpha ,\,z)$ with the axis
$z$ coinciding with the solenoid axis. In the magnetostatic case, the
solution within the infinite solenoid ($\,0 \le \,\rho \, < \,R)$ can be
chosen in the form $A\,_{1\alpha}  \,\, = \,\,c_{1} \,\rho
\,\,\,\,\,(\,A_{1}^{f} = A\,_{1\alpha}  ,\,\,\,\,A_{1}^{0} \, = \,0)${\large
.} In the outer region ($\,\rho \, > \,R)$, the solution has the form
$A\,_{2\alpha}  \,\, = \,\,c_{2} \, / \,\rho \,\, + \,c_{3} \,\rho $.{\large
}The system being infinite, one cannot require the potential to vanish at
large distances. As is clear from purely physical considerations, the
magnetic field outside the solenoid is zero, i.e., $A_{2}^{f} = \,\,0$.
Therefore, the only potential that can exist in the outer region is ${\rm
{\bf A}}^{0}$, which satisfies the additional condition $\,rot\,{\rm {\bf
A}}^{0}\, = \,\,0,$ and it is this condition that identifies the correct
solution $A\,_{2\alpha}  \,\, = \,\,c_{2} \, / \,\rho \,\,$. The potential
in the outer region is essentially the zero-field potential, so that ${\rm
{\bf A}}_{2} \, = \,grad\,\chi $, but because this region is doubly
connected, the $\chi $ function is multivalued, and ${\oint\limits_{L} {{\rm
{\bf A}}_{2} \,d{\rm {\bf l}}}} \,\, \ne \,\,0$, despite the fact that in
this region ($R\, < \,\rho \, < \,\infty )\,rot\,{\rm {\bf A}}_{2} \,
\equiv \,\,0$.{\large} This means that the Stokes theorem (in the form
${\oint\limits_{L} {{\rm {\bf A}}\,d{\rm {\bf l}}}} \,\, =
\,\,{\int\limits_{S} {{\rm {\bf B}}\,d{\rm {\bf s}}}} $, where the $L$
contour encloses the solenoid in the outer region, and $S$ is the area
bounded by contour $L)$ is inapplicable here. Nevertheless, this relation,
although invalid, is used in all studies of the AB effect.

The above separation of the potentials into the field and zero-field ones
permits one to find the zero-field potentials for a time-dependent current
as well. As before, we assume that circular currents flow in a region of
space to form an infinitely long cylinder.{\large} The reference frame will
be left unchanged. The current can be described by the following relations

 $j_{\alpha}  $ ($\rho ,\,\alpha ,\,z)${\large} $ = \,\,I_{0} \,\delta \,(\rho
\, - \,R)\,\exp \,i( - n\alpha \, + \,\omega t)\,${\large ,} $j_{\rho}  \, =
\,j_{z} \, = \,0${\large , (6)}

\noindent
where {\large $R$} is the solenoid radius, $\omega $ is the cyclic frequency of
the current, and $I_{0} \, = \,J\, / \,2\pi R$; here $J$ is the current in
the cylinder wall per unit length of the solenoid.

The nonzero vector-potential components $A_{\rho}  $ and $A_{\alpha}  $can
be written [12]

\begin{equation}
\label{eq5}
A_{\rho}  \, = \,{\int\limits_{V} {j_{\alpha}  \,(}} {\rm {\bf {\rho
}'}})\,\sin \,(\alpha \, - \,{\alpha} ')\,G\,({\rm {\bf \rho}} ,\,{\rm {\bf
{\rho} '}})\,d{V}',
\end{equation}

\begin{equation}
\label{eq6}
A_{\alpha}  \, = \,{\int\limits_{V} {j_{\alpha}  \,(}} {\rm {\bf {\rho
}'}})\,\cos \,(\alpha \, - \,{\alpha} ')\,G\,({\rm {\bf \rho}} ,\,{\rm {\bf
{\rho} '}})\,d{V}',
\end{equation}

\noindent
where $G\,({\rm {\bf \rho}} ,\,{\rm {\bf {\rho} '}})\,\, = \,\, -
\,{\frac{{i\pi}} {{c}}}\,H_{0}^{\left( {2} \right)} \,(k{\left| {\,{\rm {\bf
\rho}}  - \,{\rm {\bf {\rho} '}}} \right|})$ is the Green function of the
Helmholtz equation [12], $H_{0}^{\left( {2} \right)} $ is the Hankel
function, $k\, = \,\omega / \,c$, and $d\,{V}'\,\, = \,{\rho} '\,\,d{\rho
}'\,d\,{\alpha} '$.{\large} Here and in what follows, the harmonic
dependence on time is omitted. The integrals entering Eq. (7) can be easily
calculated using the rules of the totals for the Hankel functions [12]

\[
H_{0}^{\left( {2} \right)} \,(k\,\sqrt {\rho ^{2}\, + \,R^{2}\, - \,2\rho
R\cos \,(\alpha - {\alpha} ')} )\, = \,
\]

\begin{equation}
\label{eq7}
 = \,{\sum\limits_{m = - \infty} ^{\infty}  {e^{ - im(\alpha - \alpha '\,)}}
}
{\left\{ {{\begin{array}{*{20}c}
 {H_{m}^{\left( {2} \right)} \,(kR)\,\,J_{m} \,(k\rho ),\,\,\rho < \,R}
\hfill \\
 {J_{m} \,(kR)\,H_{m}^{\left( {2} \right)} \,(k\rho ),\,\,\rho \, > \,R}
\hfill \\
\end{array}}}  \right.}
\end{equation}

As a result, we obtain

\[
A_{\alpha}  \, = \, - \,{\frac{{i\pi ^{2}I_{0} \,R}}{{c}}}\,e^{ - in\alpha
}\,\times
\]

\begin{equation}
\label{eq8}
{\left\{ {{\begin{array}{*{20}c}
 {H_{n + 1}^{\left( {2} \right)} (kR)\,J_{n + 1} (k\rho )\, + \,H_{n -
1}^{\left( {2} \right)} \,(kR)\,J_{n - 1} (k\rho ),\,\,\rho < R} \hfill \\
 {J_{n + 1} (kR)H_{n + 1}^{\left( {2} \right)} (k\rho )\, + \,\,J_{n - 1}
(kR)H_{n - 1}^{\left( {2} \right)} \,(k\rho ),\,\,\rho > R} \hfill \\
\end{array}}}  \right.}
,
\end{equation}

\[
A_{\rho}  \, = \, - \,{\frac{{\pi ^{2}I_{0} \,R}}{{2c}}}\,e^{ - in\alpha
}\,\times
\]

 ${\left\{ {{\begin{array}{*{20}c}
 {H_{n + 1}^{\left( {2} \right)} (kR)\,J_{n + 1} (k\rho )\, - \,H_{n -
1}^{\left( {2} \right)} \,(kR)\,J_{n - 1} (k\rho ),\,\,\rho < R} \hfill \\
 {J_{n + 1} (kR)H_{n + 1}^{\left( {2} \right)} (k\rho )\, - \,\,J_{n - 1}
(kR)H_{n - 1}^{\left( {2} \right)} \,(k\rho ),\,\,\rho > R} \hfill \\
\end{array}}}  \right.}.$ (9b){\large .}

For ${\rm n} {\rm =}  {\rm 0}  $ (i.e., in the absence
of modulation in angle), Eq. (9) yield

 $A_{\alpha}  \, = \, - \,{\frac{{2i\pi ^{2}\,I_{0} \,R}}{{c}}}
{\left\{ {{\begin{array}{*{20}c}
 {H_{1}^{\left( {2} \right)} (kR)\,J_{1} \,(k\rho ),\,\rho < R} \hfill \\
 {J_{1} \,(kR)H_{1}^{\left( {2} \right)} (k\rho ),\,\rho > R} \hfill \\
\end{array}}}  \right.}$ and{\large} $A_{\rho}  \, = \,0${\large . (10)}

In the static case ($\omega \, \to \,0)$, one obtains from these relations
the well-known expressions

 $A_{\alpha}  \, = \,J\,\rho \, / \,c\,R\,\,(\rho \, < \,R)$ and{\large
}$A_{\alpha}  \, = \,J\,R\, / \,c\,\rho \,\,(\rho \, > \,R)${\large , (11)}

Consider in more detail the potential of Eq. (10) in the outer region, which
is of major interest for us here

\[
A_{\alpha}  \, = \,Q\,H_{1}^{\left( {2} \right)} (k\rho )\, \equiv \,Q
[J_{1} \,(k\rho )\, - \,\,i\,Y_{1} \,(k\rho )]
 =
\]

 $ = \,\,\,Q\,\,$ {\large {\{}}$\,{\frac{{2i}}{{\pi k\rho}} }\, + \,{\left[ {1 -
{\frac{{2iC}}{{\pi}} } - \,{\frac{{2i}}{{\pi}} }\ln \left( {{\frac{{k\rho
}}{{2}}}} \right)} \right]}{\sum\limits_{m = 0}^{\infty}  {{\frac{{( -
1)^{m}}}{{m!\,\Gamma \,(m + 2)}}}\left( {{\frac{{k\rho}} {{2}}}} \right)^{2m
+ 1}\,}} \,\, + \,$

 $ + {\frac{{i}}{{\pi}} }{\sum\limits_{m = 0}^{\infty}  {{\frac{{( -
1)^{m}}}{{m!\,(m + 1)!}}}\left( {{\frac{{k\rho}} {{2}}}} \right)^{2m + 1}}
}{\left[ {{\sum\limits_{j = 1}^{m} {{\frac{{1}}{{j}}}}} \, +
\,{\sum\limits_{j = 1}^{m + 1} {{\frac{{1}}{{j}}}}}}  \right]}$ {\large {\}},
(12)}

\noindent
where $Q\,\, = \,\, - \,{\frac{{2i\pi ^{2}\,I_{0} \,R}}{{c}}}J_{1} \,(kR)$,
$C$ is Euler's constant, and $Y_{1} $ is the Neumann function.

As seen from Eq. (12), the curl of the first term in braces is zero. One can
readily verify that the curls of the other terms in the braces are nonzero.
Thus, in this case the total potential can be separated into the field and
the zero-field potential. As follows from Eq. (1)

\begin{equation}
\label{eq9}
\varphi ^{0}\, =
\quad
{\left\{ {{\begin{array}{*{20}c}

{0,\,\,\,\,\,\,\,\,\,\,\,\,\,\,\,\,\,\,\,\,\,\,\,\,\,\,\,\,\,\,\,\,\,\,\,\,\,\,\,\,\,\,\,\,\,\,\,\rho
< R} \hfill \\
 { - \,{\frac{{4\pi iI_{0} \,R}}{{c}}}\,J_{1} \,(kR)\,\alpha
,\,\,\,\,\,\,\,\rho > R} \hfill \\
\end{array}}}  \right.} \quad .
\end{equation}

Separation of the real part of the components of the potentials in Eq. (12)
yields [13, 14]

{\large Re}$\,A_{\alpha} ^{f} \, = \,W\,\{\pi J_{1} (k\rho )\,\sin \omega
t\, - \,{\left[ {{\frac{{2}}{{k\rho}} }\, + \,\pi Y_{1} (k\rho )}
\right]}\,\cos \omega t\}${\large , (14a)}

{\large Re}$A_{\alpha} ^{0} \, = \,W\,{\frac{{2}}{{k\rho}} }\,\cos \omega
t${\large , (14b)}

\noindent
where $W\, = \,{\frac{{2\pi I_{0} \,R\,J_{1} (kR)}}{{c}}}$.

One can readily calculate the flux inside the solenoid

\begin{equation}
\label{eq10}
\Phi \, = \,{\int\limits_{0}^{R} {{\int\limits_{0}^{2\pi}  {B_{z} \,(\rho
)\,\rho \,d\rho \,d\alpha \, = \, - \,{\frac{{i\,4\,\pi ^{3}\,R^{2}\,I_{0}
}}{{c}}}}}} } \,J_{1} \,(kR)\,H_{1}^{\left( {2} \right)} \,(kR)
\end{equation}

At the same time, the circulation of vector ${\rm {\bf A}}^{0}$ along an
arbitrary contour enclosing the solenoid (the cyclic constant)

\begin{equation}
\label{eq11}
\omega _{1} \, = \,\,{\oint\limits_{L} {{\rm {\bf A}}^{0}}} d{\rm {\bf l}}\,
= \,{\frac{{8\pi ^{3}\,I_{0} \,R}}{{ck}}}J_{1} \,(kR).
\end{equation}

Obviously enough, in a general (time-dependent) case $\omega _{1} \, \ne
\,\Phi $, and therefore their coincidence in the static case ($k \to \,0)$
should be accepted as purely accidental.

Consider now the geometry of the Aharonov-Bohm experiment, in which
electrons move around a solenoid along a circle of a given radius.{\large
}We shall limit ourselves to the case where the electrons meet on their way
nonzero zero-field potentials, while field potentials are not present. This
situation can be realized by enclosing the solenoid in cylindrical screens,
or, as follows from Eq. (13a),{\large} by choosing the trajectory radii of
the electrons and by mathching properly their transit with the current
variation in the solenoid.{\large} Substituting now the zero-field
potentials in the Schrodinger equation and using the procedure of the
solution proposed in (Appendices B and D in [10]) but, in contrast to [10],
performing time averaging, we come to the following expression for the
intensity of the interference pattern [15]

\begin{equation}
\label{eq12}
\overline {P} = \,0.5\,\,P_{0} \,\,\,{\left\{ {1 + \,J_{0} \,(S) \cdot \cos
\,[\omega _{e} \,\tau ]} \right\}},
\end{equation}

\noindent
where $S\,\, = \,\,16\,\pi ^{3}\,I_{0} \,\,R\,\mu _{0}^{ - 1} \,\omega ^{ -
1}\,J_{1} \left( {k\,R} \right)$, $\mu \,_{0} \,\,\, =
\,\,{\raise0.7ex\hbox{${ch}$} \!\mathord{\left/ {\vphantom {{ch} {{\left|
{e} \right|}}}}\right.\kern-\nulldelimiterspace}\!\lower0.7ex\hbox{${{\left|
{e} \right|}}$}}\,$ and $J_{0} \,\,$ and $J_{1} $ are the Bessel functions.
For $I_{0} \,\, = \,\,\,158\,\,mA / cm\,\,;\,\,R\,\, = \,\,5\,\mu
m\,;\,\,\omega / 2\pi \,\, < \,\,10^{10}Hz$, we obtain $S$ = 2.45.{\large} This
means that the interference pattern should vanish for these parameters. To
verify experimentally this conclusion, one should use preferably electrons
in metallic mesoscopic rings or cylinders [4, 5]

Thus, we believe that the Aharonov-Bohm experiment in both the static and
the time-dependent case is actually an experiment on detection of a field of
a new type in classical electrodynamics. This field has none of the
characteristics inherent in the classical electromagnetic fields, namely,
the energy, the momentum, and the angular momentum. Therefore, these fields
have a high penetration capacity and can be used for information transfer,
with its detection by the AB effect.

REFERENCES

{\large 1. W. Franz, Verhandlungen der Deutschen Physikalischen
Gesellschaft. 20, 65 (1939).}

{\large 2. W. Ehrenberg and R. E. Siday, Proc. Phys. Soc. London, Sect. Â
62, 8 (1949).}

{\large 3. Y. Aharonov and D. Bohm, Phys. Rev. 115, 485 (1959).}

{\large 4. S. Olariu and I. I. Popesku, Rev. Mod. Phys. 57, 339 (1985).}

{\large 5. M. Peskin and A. Tonomura, Lect. Notes Phys. 340, 115 (1989).}

{\large 6. R. D. Richtmyer,} {\large {\it Principles of Advanced Mathematical Physics}} {\large (Springer-Verlag? New York,
1978), Vol. 1-2; B. A. Dubrovin, S. P. Novicov, A. T. Fomenco,} {\large
{\it Modern geometry}}{\large (Moscow, 1986).}

{\large 7. L. H. Ryder,} {\large {\it Quantum Field Theory}}{\large (Cambridge University Press,
Cambridge, London, 1985).}

{\large 8. R. G. Chambers, Phys. Rev. Lett. 5, 3 (1960).}

{\large 9. A. Tonomura et al., Phys. Rev. Lett. 56, 792 (1986).}

{\large 10. B. Lee, E. Yin, Ò. Ê. Gustafson, and R. Chiao, Phys. Rev. A 45,
4319 (1992).}

{\large 11. A. N. Tikhonov and A. A. Samarskii, Equations of Mathematical
Physics (Nauka, Moscow, 1953; Pergamon, Oxford, 1964).}

{\large 12. G. T. Markov and A. F. Chaplin, Excitation of Electromagnetic
Waves (Moscow, 1967).}

{\large 13. A. G. Chirkov and A. N. Ageev, Pis'ma Zh. Tekh. Fiz. 26, 103
(2000) [Tech. Phys. Lett. 26, 747 (2000)].}

{\large 14. A. G. Chirkov and A. N. Ageev, Technical Physics, 46, 147
(2001).}

{\large 15. A. G. Chirkov and A. N. Ageev, Solid State Physics (in
press).}

\end{document}